\documentclass[sigconf, nonacm, preprint]{acmart}
\usepackage{graphicx} %
\usepackage{enumitem}
\usepackage{subcaption}
\usepackage{tabularx}
\usepackage{booktabs}

\title[Shift-Up: Framework for Software Engineering Guardrails in AI-native Software Development - Initial findings]{Shift-Up: A Framework for Software Engineering Guardrails in AI-native Software Development - Initial Findings}

\author{Petrus Lipsanen}
\authornote{Both authors contributed equally to this research.}
\affiliation{%
  \institution{University of Jyväskylä}
  \city{Jyväskylä}
  \country{Finland}}
\email{petrus.i.lipsanen@jyu.fi}

\author{Liisa Rannikko}
\authornotemark[1]
\affiliation{%
  \institution{University of Jyväskylä}
  \city{Jyväskylä}
  \country{Finland}}
\email{liisa.m.e.rannikko@jyu.fi}

\author{François Christophe}
\affiliation{%
  \institution{University of Jyväskylä}
  \city{Jyväskylä}
  \country{Finland}}
\email{francois.m.christophe@jyu.fi}

\author{Konsta Kalliokoski}
\affiliation{%
  \institution{University of Jyväskylä}
  \city{Jyväskylä}
  \country{Finland}}
\email{konsta.m.kalliokoski@jyu.fi}

\author{Vlad Stirbu}
\affiliation{%
  \institution{University of Jyväskylä}
  \city{Jyväskylä}
  \country{Finland}}
\email{vlad.a.stirbu@jyu.fi}

\author{Tommi Mikkonen}
\affiliation{%
  \institution{University of Jyväskylä}
  \city{Jyväskylä}
  \country{Finland}}
\email{tommi.j.mikkonen@jyu.fi}

\date{March 2026}

\begin{document}

\acmConference[EASE 2026]{The 30th International Conference on Evaluation and Assessment in Software Engineering}{9–12 June, 2026}{Glasgow, Scotland, United Kingdom}
\acmYear{2026}
\copyrightyear{2026}

\begin{abstract}
Generative AI (GenAI) is reshaping software engineering by shifting development from manual coding toward agent-driven implementation. While “vibe coding” promises rapid prototyping, it often suffers from architectural drift, limited traceability, and reduced maintainability. Applying the design science research (DSR) methodology, this paper proposes Shift-Up, a framework that reinterprets established software engineering practices, like executable requirements (BDD), architectural modeling (C4), and architecture decision records (ADRs), as structural guardrails for GenAI-native development. Preliminary findings from our exploratory evaluation compare unstructured vibe coding, structured prompt engineering, and the Shift-Up approach in the development of a web application. These findings indicate that embedding machine-readable requirements and architectural artifacts stabilizes agent behavior, reduces implementation drift, and shifts human effort toward higher-level design and validation activities. The results suggest that traditional software engineering artifacts can serve as effective control mechanisms in AI-assisted development.
\end{abstract}


\maketitle

\section{Introduction}

The emergence of Generative AI (GenAI) has had a profound impact on software engineering \cite{belzner2023large}. GenAI is becoming increasingly capable of aiding different software engineering tasks, whether it be requirements engineering, code generation, or documentation \cite{nguyen2024generative}. This has led to their widespread adoption in software development processes \cite{stray2025generative}, which in turn has created a shift in which actual software development is performed by GenAI development environments such as {Lovable}\footnote{\href{https://lovable.dev/}{https://lovable.dev/, accessed on 25.2.2026}},
{Replit}\footnote{\href{https://replit.com/}{https://replit.com/, accessed on 25.2.2026}} or
{Bolt.new}\footnote{\href{https://bolt.new/}{https://bolt.new/, accessed on 25.2.2026}}.

In such environment, the activity of developers focus on prompting user needs for the software application to be developed, evaluating the functionality of the AI-generated prototype, and iterating this prompting process until the generated prototype reaches some kind of qualitative acceptance from developers. This emergent software development process is widely known as vibe coding \cite{ge_survey_2025}. The main promises of vibe coding environment rely on acceleration of delivery and software development expertise not being necessary. However, a recent study shows that these promises are often given at the cost of quality of prototypes, maintainability and rework iterations before achieving the status of a viable product \cite{pattyn_vibe_2026}. This gap needs attention because of the conflict between development acceleration and end-product quality.

To address this gap, our paper proposes Shift-Up, a framework for GenAI-native software development that leverages established software engineering practices as structural guardrails to guide and stabilize agent-driven implementation, embedding machine-readable and traceable artifacts directly into the development workflow. Through this investigation, we provide initial design knowledge on how traditional software engineering practices can be reinterpreted to structure and control AI-assisted development processes.

\section{Background and motivation}
\label{sec:background}

Software engineering has repeatedly undergone paradigm shifts that initially deprioritize established prescriptive design knowledge in favor of speed and flexibility. The transition from waterfall to agile development is a notable example. Early agile adoption emphasized working software over documentation, abandoning formal prescriptions. However, the resulting structural deficits forced the reintroduction of this design knowledge in lightweight forms such as test-driven development (TDD) and continuous integration (CI).

Several established frameworks function as prescriptive design knowledge. The V-model \cite{vmodel} operationalizes design control in quality-sensitive domains \cite{nooper_secure_2022}. Shift-left and shift-right principles \cite{shift-left,shift-right} prescribe continuous validation throughout the life cycle. Architectural modeling approaches such as the C4 model \cite{c4} and architecture decision records (ADRs) \cite{nygard2011adr,zimmermann2015adr} provide formal prescriptions for preserving system representations and design rationale. Furthermore, executable requirements \cite{fuchs99attemptoControlled}, including behavior-driven development (BDD) \cite{solis2011bdd}, constitute prescriptive knowledge that dictates how stakeholder intent must link to automated validation.

A comparable dynamic appears to be emerging in the context of GenAI-assisted development. Agent-driven workflows and vibe coding emphasize rapid prototyping through iterative prompting, often suggesting that traditional engineering artifacts are unnecessary. However, recent studies report recurring issues such as architectural drift, limited traceability, and reduced controllability in purely prompt-driven approaches \cite{russo2024navigating,mikkonen2025software,russo2024generative}. This mirrors earlier transitions in software engineering, where the temporary abandonment of structure was followed by the reintroduction of practices in adapted forms.

This historical pattern motivates our investigation: rather than discarding established software engineering practices in AI-native development, how can they be reinterpreted as lightweight structural guardrails compatible with agent-driven workflows?

\section{Methodology and objectives}

This work adopts the Design Science Research (DSR) paradigm to investigate how established software engineering practices can be reinterpreted as structural guardrails in GenAI-native development. DSR provides a rigorous, utility-driven process for designing, developing and evaluating information technology (IT) artifacts to solve complex practical problems \cite{dsr2020hevner}. The use of DSR ensures a structured pathway aligning with DSR methodologies that emphasize continuous context-aware artifact evolution \cite{dsr2024baustein, dsr2024tuunanen}.

Following the DSR process model, the central problem addressed in this evaluation is the architectural drift, limited traceability, and reduced controllability observed in unstructured agentic workflows during GenAI-native development. The solution objectives focus on establishing structural guardrails that constrain and stabilize generative behavior. The Shift-Up framework constitutes the proposed artifact, which integrates structured requirements engineering (BDD-based executable requirements), architectural modeling (C4), and architecture decision records (ADRs) into GenAI-assisted development. These artifacts are designed not only as human-readable documentation, but as machine-readable, persistent contextual constraints.

The demonstration of this framework consists of applying the structured artifacts (BDD executables, C4, ADRs) as continuous inputs within the development pipeline. Following this, the present evaluation represents the initial exploratory evaluation of this framework. The objective is to assess the feasibility of guardrail mechanisms and refine the conceptual understanding of structured artifacts in agentic workflows. This objective is distilled in the following research questions:

\begin{description}[nosep]
    \item [RQ1:] How does the use of structured requirements and architectural artifacts (ADR, C4, BDD) influence the autonomy of GenAI agents during implementation?
    \item [RQ2:] To what extent do executable requirements reduce agent drift compared to prompt-only development?\
\end{description}
Together, these questions aim to uncover the mechanisms through which traditional software engineering artifacts \cite{nooper_secure_2022} can serve as structural stabilizers in GenAI-native development processes.

\section{The Shift-Up framework}

\begin{figure}
    \centering
    \includegraphics[width=0.95\linewidth]{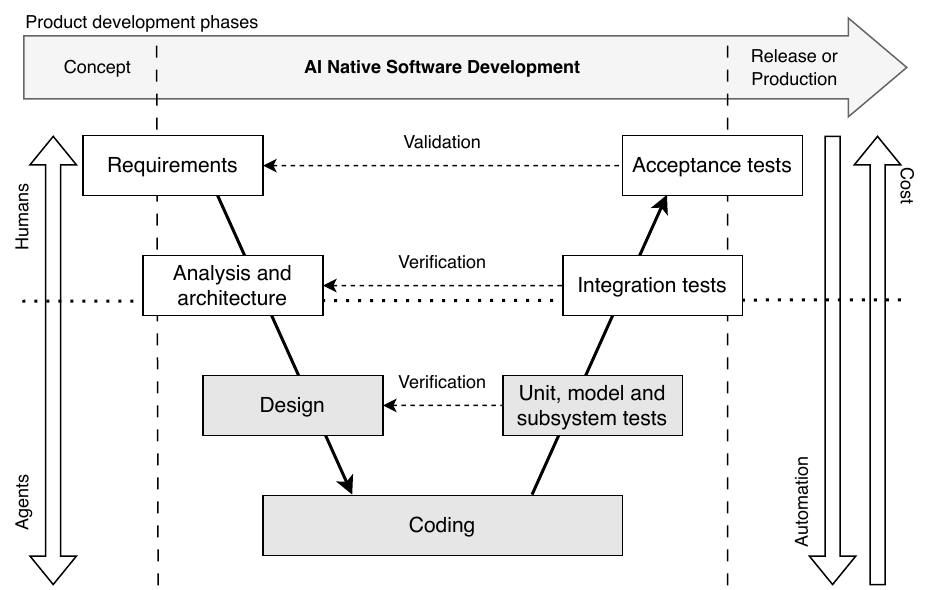}
    \caption{AI-native software development with Shift-Up}
    \Description{Diagram shoeing Ai-native development with Shift-Up.}
    \label{fig:ai-native}
\end{figure}

Assisted by GenAI tools, development teams can generate artifacts across multiple levels of abstraction from small code snippets to fully deployable modules integrated and monitored into larger systems. In this setting, shift-left acquires a renewed meaning: the emphasis moves from manual coding to higher-level artifacts such as prompts, refined specifications, and automated generation, placing developer intent at the center of the life cycle. In addition, shift-right extends beyond traditional testing toward deployment and operations, where GenAI supports continuous monitoring, validation, and feedback-driven evolution to ensure robustness under production conditions.

By combining these perspectives, we propose the concept of \textit{Shift-Up}, which fits the layered structure of the V-Model. In this GenAI-native Software Development Life-Cycle (SDLC), intermediate stages such as detailed design, implementation, and low-level verification are delegated to GenAI tools, allowing human developers to focus on the higher layers of the V-Model. At the top-left, they concentrate on requirement specification, architectural design, and system-level considerations, and at the top-right, they engage in acceptance testing, deployment oversight, and operational feedback. The objective of Shift-Up is thus to free developers from low-level implementation details and empower them to focus on strategic, creative, and domain-specific aspects of software development, while ensuring that quality and reliability are preserved throughout the life cycle, as depicted in Figure \ref{fig:ai-native}. Upon performing Shift-Up development, humans and GenAI co-operate to solve problems in ways that are best fit to the task at hand, following a shared responsibility working model \cite{se-bots}.

\section{Exploratory comparative evaluation: Shift-Up framework in practice} \label{sec:research}

To evaluate the \textit{Shift-Up} framework, we conducted a comparative evaluation centered on the development of a web application for an online snack-bar. The scope of this snack-bar is a full-stack application that includes graphical user interface (GUI), PostgreSQL database, admin system, and other backend logic. The evaluation consists of three distinct developmental paradigms:
\begin{enumerate}[nosep]
    \item \textbf{Unstructured vibe coding:} An informal development approach in which implementation progresses directly from high-level ideas, with limited upfront requirements modeling, architectural planning, or formal validation artifacts. This is referred in the industry as "vibe coding".
    \item \textbf{Structured vibe coding through prompt engineering:} This approach prioritizes an iterative process characterized by reliance on agentic AI with minimal formal scaffolding achieved by prompt engineering.
    \item \textbf{Initial experimentation of Shift-Up:} Partial application of the proposed framework specifically focusing on the initial two layers: requirements engineering and automated validation through executable acceptance tests, with agent driven architecture design.
\end{enumerate}

The objective is to analyze how structured requirements, combined with executable acceptance tests and generated C4 and ADR artifacts, influence the reliability and validity of AI-generated software and the development process. The methodology is then compared to a purely prompt-driven approach. The authors worked as human developers, one author working through a single approach. In none of the approaches did a human write any code; the responsibility was delegated to the GenAI agent through prompting. 

\textbf{Data collection}. Each prompt was recorded alongside a qualitative assessment of its efficacy. To capture the developer experience, developers were to document notes and observations in a journal. Quantitative metrics such as implementation time were recorded for all approaches. As this is an exploratory comparative evaluation, some prompt practices were also iteratively modified during the development process. A subsequent analysis will focus on the technical evaluation of the generated code and test suites. The current results focus on the implementation process.

\textbf{Data analysis}. The analysis of the prompts and coding of the categories was aligned with a general inductive approach \cite{inductiveThomas2006}. The authors working on structured vibe coding and Shift-Up approaches re-familiarized themselves with, analyzed, and categorized their own prompts. Subsequently, the authors cross-validated each other’s categories to ensure consistency. The inductive approach was chosen due to the exploratory nature of this evaluation.

\subsection{Unstructured vibe coding}
\textbf{Unstructured vibe coding} \footnote{\href{https://github.com/Lovablekokeilu/campus-treats}{Unstructured vibe coding GitHub repository}} was done via the Lovable app, which promises that there is no need for coding proficiency. The approach started through a short semi-structured interview with the stakeholder. The aim for the interview was for the developer to gain a vision for the future application and key characteristics of the functionality. Afterwards the developer made a description of the application and prompted the Lovable system through its GUI. 

The system worked well for generating GUI and quickly hosting it in the Lovable cloud system in a couple of prompts. Although the system works well for these kind of application, the scope is very limited. As Lovable has their own systems/guardrails in place it limits the choices of the developer. This also forces the developer to switch platforms and tools if they  later want to host it somewhere else. Due to these reasons, this approach was left on the prototyping level and out of the larger comparison. 

\subsection{Structured vibe coding: Prompt engineering approach}
The \textbf{structured vibe coding} \footnote{\href{https://github.com/JYU-GENIUS-project/VibeCode1}{Structured vibe coding GitHub repository}} approach started through an interview with the stakeholder which was the same as for the unstructured vibe coding along with the notes gathered. Next, the developer generated an implementation plan based on the notes collected from the interview and started working from there. During the implementation, the vibe developer used some prompting strategies (e.g. planning, progress description files) to manage the process and context. The project was done utilising VS Code and GPT-5.0-Codex agent.

The first prompt was to plan an implementation for the whole process. The plans generated during implementation also included checklists to manage context and to keep the developer up to date on the progress. The developer used plans according to their view on when a plan is needed for a larger implementation and when only a prompt would suffice for a more minor functionality. 

The original plan generated in the first prompt ended up guiding the agentic development to a skeleton for the product. After this, the developer guided the agent on what kind of functionality the developer thought needed to be implemented. The development cycle was mostly to prompt the agent to implement a feature and then manually testing if it worked. The developer never prompted the agent to generate unit tests but some were generated and proved useful in identifying some breaking changes during development.

\subsection{Shift-Up approach}

\begin{figure*}[ht]
    \centering
    \begin{subfigure}[b]{0.64\textwidth}
        \centering
        \includegraphics[width=\linewidth]{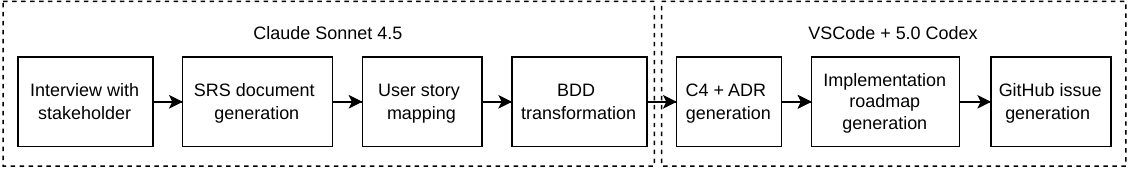}
        \caption{}
        \label{fig:shiftup1}
    \end{subfigure}\hfill
    \begin{subfigure}[b]{0.355\textwidth}
        \centering
        \includegraphics[width=\linewidth]{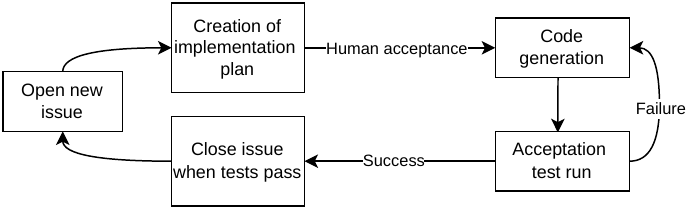}
        \caption{}
        \label{fig:shiftup2}
    \end{subfigure}
    
    \caption{Shift-Up workflow: (a) requirements and architectural grounding, and (b) GenAI-assisted implementation.} %
    \label{fig:shiftup}
    \Description{A figure describing the Shift-Up workflow in two parts: (a) requirements and architectural grounding workflow, and (b) GenAI-assisted implementation cycle.}
\end{figure*}

The \textbf{Shift-Up} framework \footnote{\href{https://github.com/JYU-GENIUS-project/snackbar_v1}{Shift-Up approach GitHub repository}} was used to transform stakeholder needs into a structured machine-readable context. This context is then used first as guardrails for code generation and second as acceptance tests for the generated code.

The first part of this framework, shown in Figure~\ref{fig:shiftup1}, proceeded with the following steps: \textbf{Stakeholder interview} -- the experimental process began with a more in-depth stakeholder interview than in previous processes. To further elicit core functional and non-functional requirements, Claude Sonnet 4.5 was used as an interactive assistant; \textbf{SRS generation} -- the interview output was synthesized into a Software Requirements Specification (SRS) document. The SRS was then refined through three iteration rounds, each guided by prompt: \textit{"Identify any ambiguous or vague terms and flag any requirements that seem to contradict"}; \textbf{User story mapping} -- the refined SRS were further decomposed into user stories with the prompt: \textit{"Provide structured user stories based on the SRS document. Use a template 'As a <user>, I want <goal>, so that <benefit>'"}. In total 68 user stories were created; \textbf{BDD transformation} -- each user story was decomposed into executable acceptance tests in Robot Framework \footnote{\href{https://robotframework.org/}{Robot Framework, accessed on 26.2.2026}} format with the prompt: \textit{"Write an acceptance test-case for each user story in Robot Framework format. Write them in given-when-then format and use best practices"}. In total 175 test cases were generated; \textbf{Generation of C4 and ADR artifacts} -- using the previous steps, C4 and ADR artifacts were generated for context purposes; \textbf{Implementation roadmap generation} -- the project was composed into a roadmap where features were divided into implementation phases with the prompt \textit{"Make a comprehensive step-by-step implementation plan for this project. Break it down into sequential batches based on themes of the acceptance tests and logical dependencies. Format: name, goal/theme, key architecture tasks, acceptance test IDs to validate, dependencies"}. This produced 10 implementation phases that were ordered by logical dependency; finally, \textbf{GitHub issue generation} -- structured GitHub issues were created based on implementation phases in the roadmap with the prompt \textit{"Create issues for each phase in roadmap. Format: implement [name], description, constraint (test id's that must pass for issue to close), context, links"}. The first four phases were done utilizing Claude Sonnet 4.5 while the last three were completed using GPT-5.0-Codex integrated within VS Code.

The implementation and verification loop was the second part of the framework, shown in Figure~\ref{fig:shiftup2}, which proceeded one phase at a time. Each phase was developed in VS Code integrated development environment (IDE) and GPT-5.0-Codex in its own Git branch and submitted as a pull request upon completion. Each loop consisted of five steps: \textbf{Open issue} -- the developer selects the next GitHub Issue from the roadmap, respecting the dependencies established in the plan phase; \textbf{Implementation plan creation} -- implementation begins with the standardized prompt, which was iteratively refined during the evaluation: \textit{"Create an implementation plan for features described in issue X, strictly adhering to architectural plans and technical requirements found in the workspace. Break the implementation down into logical, sequential subtasks and identify which acceptance criteria/tests apply to each phase. Create a separate .md file that includes the plan"}; \textbf{Code Generation} -- after the plan has been established, the agent proceeds with implementation. In practice, after the plan was created and reviewed, the only instruction given by the developer was to proceed with the plan. Afterwards, the agent generates source code, configuration, and supporting files according to the sequential subtasks in its plan; \textbf{Acceptance test execution} -- the Robot Framework acceptance tests linked to the current issue are executed. Two outcomes are possible. Firstly, if all linked tests pass, we simply go to the next step. Secondly, if one or more tests fail, the agent receives the output of these tests as an additional context for the additional iteration of the code generation step. This cycle repeats until all constrained acceptance tests have passed; \textbf{Close issue} -- after all tests have passed, the open issue is closed, and the developer opens a new issue.

\section{Results}
\label{sec:disc}

\begin{table*}
    \centering
    \caption{Qualitative evaluation of the approaches covered in exploratory comparative evaluation}
    \label{tab:eval-explor-study}
    \begin{tabular}{lccccc}
        \toprule
         Development Paradigms&  Upfront Investment&  Human Control&  Structured Constraints&  Development Speed& Guardrails\\
         \midrule
         Unstructured & Minimal & Low & - & Fastest prototype & -\\
         Structured& Minimal & Moderate & Structured & Fast & Prompt Engineering\\
         Shift-Up& High & High & Rigid & Slower & BDD/TDD, C4, ADR\\
         \bottomrule
    \end{tabular}
\end{table*}

In total 176 recorded prompts were categorized from each the structured vibe coding and the Shift-Up approach implementation phases. These are preliminary results, as the Shift-Up approach implementation is not yet concluded, but they provide an overview of the prompt patterns that emerged from the experiment so far.

In the Shift-Up approach, the interactions focused primarily on process orchestration and automated validation. Thematic distribution of the prompts revealed five distinct categories: \textit{proceeding with the next step} (62~\%), \textit{executing acceptance tests} (16~\%), \textit{developer identified fixes} (9~\%), \textit{acceptance of agent-proposed solutions} (7~\%), and \textit{initiating the next step of the overall plan} (5~\%). In contrast, the structured vibe coding approach was characterized by the developer reacting to agent output. More than half of all prompts were dedicated to \textit{addressing issues identified manually in GUI or IDE} (52~\%), followed by \textit{proceeding with the next step} (27~\%). Strategic prompts such as \textit{initial planning of a feature} and \textit{new feature implementation} each accounted for 5~\% of the total distribution and 11 \% of prompts were categorized as other.

This thematic categorization revealed a divergence in the developer interaction: the Shift-Up approach is characterized by strategic orchestration, with the majority of the prompts dedicated to advancing the implementation and verifying via automated test suites. In contrast, the structured vibe coding consists of reactive intervention, where the role of the developer is to identify and remediate issues surfaced within the GUI or IDE. Regarding RQ1, we found that by establishing a foundation of autonomous continuous validation, the generative agent is enabled to operate with increased independence during implementation tasks. The integration of BDD principles and executable requirements function as behavioral guardrails allowing the agent to self-validate its progress.

Table 1 presents a qualitative evaluation of the approaches according to 5 different categories: upfront investment, human control, structured constraints, development speed, and guardrails. This evaluation compares between the approaches and focuses on the themes from the viewpoint of the developer. Upfront investment was evaluated as the workload prior to starting implementation, human control was evaluated as the ability of the developer to guide the agent during implementation, structured constraints were evaluated as the formal rules of the approach or limitations imposed on design and code of the system, development speed was evaluated as the time to implement an idea, and guardrails are what were used to enforce human control beyond prompts.

The main result from this qualitative evaluation is that Shift-Up has more human control, structured constraints, and guardrails in exchange for higher workload in upfront investment and slower development speed. The structured and unstructured vibe coding approaches were similar in upfront investment, but differences emerged from the amount of human control. For the unstructured approach, the exchange was giving control to the system and locking some solutions in place for a quickly developed prototype.

While the Shift-Up framework constrains agent behavior, we were only able to partially answer RQ2 regarding the reduction of agent drift. As an exploratory evaluation, it was intentionally limited to a small project in a relatively common application domain. This likely made the project more feasible with a vibe coding approach, but this commonality could be a reason for preventing us from observing drift. From this comparative evaluation, we hypothesize that to evaluate agentic drift, the target application needs a more complex domain of application.

\section{Discussion and conclusion} %
\label{sec:conclusions}

This paper introduced and explored Shift-Up, formalizing it as prescriptive design knowledge for GenAI-native software engineering. By recontextualizing established practices, specifically C4 modeling, architecture decision records, and behavior-driven development, as structural and behavioral guardrails, Shift-Up transitions the operational scope of the developer from low-level code authorship to high-level system orchestration.

A primary theoretical contribution is the ability of the selected guardrails to enhance agent autonomy, which leads to the redistribution of developer attention throughout the development. In the structured vibe coding approach, the developer effort was concentrated on reactive debugging and manual verification. In contrast, Shift-Up necessitates greater engagement during the initial requirements engineering and design phases. The key tradeoff that achieved this greater agent autonomy was the increased workload prior to implementation phase.

The Shift-Up approach demonstrates a new paradigm for GenAI-native software development. By supplying the developer with deterministic prescriptive guardrails rather than relying on probabilistic prompt optimization, the framework proves that future autonomous AI-native systems must remain grounded in established software engineering design knowledge.

\begin{acks}
This evaluation was supported by Business Finland projects ITEA GENIUS (2545/31/2024) and ANSE (1822/31/2025).
\end{acks}

\bibliographystyle{acm}
\bibliography{bibliography, refs}

\end{document}